\documentstyle[aps,multicol]{revtex}

\renewcommand{\narrowtext}{\begin{multicols}{2} \global\columnwidth20.5pc}
\renewcommand{\widetext}{\end{multicols} \global\columnwidth42.5pc}
\newfont{\abg}{cmsl12} 

\def\nn{\nonumber \\}

\def\be{\begin{equation}}
\def\ee{\end{equation}}
\def\ba{\begin{eqnarray}}
\def\ea{\end{eqnarray}}
\def\pl{\label} 
\def\re{(\ref }
\def\rz#1 {(\ref{#1}) }   \def\ry#1 {(\ref{#1})}
\def\el#1 {\label{#1}\end{equation}}

\let\a=\alpha \let\b=\beta \let\g=\gamma \let\d=\delta
\let\e=\varepsilon \let\ep=\epsilon   
   \let\m=\mu
    \let\s=\sigma
 \let\o=\omega

\def\0{\over} \def\1{\vec} \def\2{{1\over2}} \def\4{{1\over4}}
\def\5{\bar} \def\6{\partial}

\def\({\left(} \def\){\right)} \def\<{\langle} \def\>{\rangle}
 
\def\[{\lbrack} \def\]{\rbrack}

\def\wt{\widetilde}
 \let\ra=\rightarrow

         \def\CM{{\cal M}} 

 \def\CP{{\cal P}}

\def\dl{\overleftarrow{\6}}
\def\dr{\overrightarrow{\6}}
 
\begin{document}

\title{Target--superspace in 2d dilatonic supergravity}

\author{Thomas Strobl}

\address{Institut f\"ur Theoretische Physik, RWTH Aachen, D-52056
Aachen, Germany}

\date{April 1999}

\maketitle

{\tightenlines
\begin{abstract}
  The $N=1$ supersymmetric version of generalized 2d dilaton gravity
  can be cast into the form of a Poisson $\s$--model, where the target
  space and its Poisson bracket are graded. The target space consists
  of a 1+1 superspace and the dilaton, which is the generator of
  Lorentz boosts therein.  The Poisson bracket on the target space
  induces the invariance of the worldsheet theory against both
  diffeomorphisms and local supersymmetry transformations
  (superdiffeomorphisms). The machinery of Poisson $\s$--models is
  then used to find the general local solution to the field
  equations. As a byproduct, classical equivalence between the bosonic
  theory and its supersymmetric extension is found. 
\end{abstract}
}

\narrowtext 
The most general 2d gravity action for a metric $g$ and a dilaton
field $\phi$ yielding second order differential equations is of the
form \cite{Banks}: \be L= \int_{\CM} d^2x \sqrt{- g} \; [U(\phi) R+
V(\phi) (\nabla \phi)^2 +W(\phi) ] \, , \el gdil where $R$ denotes the
Ricci scalar and $U$, $V$, and $W$ are some arbitrary (reasonable)
functions of the dilaton. Its supersymmetrization, considered first in
\cite{Strominger}, may be obtained in a most straightforward manner by
using the superfield formalism of \cite{Howe}. In this framework the
action takes the same form as above, where, however, each term is
replaced by an appropriately constrained supersymmetric extension and,
simultaneously, the volume form $d^2x$ is replaced by its ({\em
  worldsheet}) superspace analog $d^2x \, d^2\theta$. A term such as
$U(\phi)$, e.g., is replaced by $U(\Phi)$, where $\Phi$ is the
superfield $\Phi = \phi + i \bar \theta \xi +i \bar \theta \theta f$
with $\phi$ being the bosonic dilaton field, $\xi$ a Majorana
spinorial superpartner, and $f$ an auxiliary bosonic scalar. (For
further details the reader is referred to the literature cited above).

For many practical calculations it is necessary to reexpress the
supersymmetric extension of $L$ in terms of its component fields
($\phi$, $\xi$, $f$ etc.). The resulting action and all the more its
field equations become lengthy and their analysis involved. In the
present letter we propose a different formulation, which greatly
simplifies not only the notation but also the analysis of the
supersymmetric theory.

This latter formulation is provided by a Poisson $\sigma$--model
\cite{PSM}, the definition of which we will briefly recapitulate now,
generalizing it to the case of graded Poisson manifolds, before we
then come to its relation to \re{gdil}) (for an introduction to
bosonic Poisson $\s$--models cf.\ \cite{LNP}). The action is a
functional of $n$ scalar fields $X^i$, $i = 1, \ldots, n$, on the one
hand, which may be viewed as coordinates in an $n$--dimensional (not
necessarily linear) manifold $N$, as well as of $n$ oneforms $A_i
\equiv A_{\m i} dx^\m$, on the other hand, which may be regarded as
oneforms on the worldsheet (coordinates $x^\m$, $\m = 0,1$) taking
values in $T^* N$ (more precisely, $(A_i)$ is a oneform on the world sheet
which is simultaneously the pullback of a section of $T^* N$ by the
map $X(x)$).  In the present context we allow $N$ to carry also a
$Z_2$--grading, i.e.\ some of the fields $X^i$ and $A_{\m i}$ may be
Grassmann valued, $\s_i$ denoting their respective parity (so $X^i X^j
= (-1)^{\s_i\s_j} X^jX^i$ etc.).

To define an action we require $N$ to be equipped with a (graded)
Poisson bracket \be \{ X^i , X^j \} \equiv \CP^{ij} \, . \label{bracket}
\ee Note that the Poisson bracket is anti--symmetric only for the case
that at least one of its entries is an even (commuting) quantity; in
general one has $\CP^{ij}=(-1)^{\s_i\s_j+1} \CP^{ji}$ (while
$\CP^{ij}$ itself has grading $\s_i + \s_j$).  In terms of the 
two--tensor $\CP^{ij}$ the
standard, graded Jacobi identity (cf, e.g., \cite{Henneaux}) may
be brought into the form \be (-1)^{\s_i\s_k} \( \CP^{ij}
\frac{\dl}{\partial X^s} \) \CP^{sk} + cycl(ijk) = 0 \,
, \pl{Jacobi} \ee where a sum over the index $s$ is
understood (but not over $i$ or $k$ in the first of the three
cyclic terms). Using left derivatives $\dr = \partial$ (in
contrast to the above right derivatives $\dl$), this equation may be
written equivalently as $(-1)^{\s_i\s_k}\CP^{is} \partial_s
\CP^{jk} + cycl(ijk) = 0$.

The action of the 2d theory is \be S = \int A_i \wedge dX^i - \2 A_i
\wedge A_j \, \CP^{ji} \, , \label{PSM}\ee where the order of the
terms and indices has been chosen so as to avoid unnecessary signs in
the considerations to follow while simultaneously $S$ coincides with
its purely bosonic counterpart in the previous literature.

Before further investigating the general model \re{PSM}), we now
discuss its relation to $N=1$ supersymmetric dilaton gravity. For this
purpose we first restrict our attention to the case where $U \equiv
\phi$ and $V \equiv 0$ in \re{gdil}). This is not a serious
restriction since the general action can be brought into this form
always (at least locally) by a dilaton--dependent conformal rescaling
of the metric $g$ and a simultaneous change of the dilaton field $\phi
\ra F(\phi)$ \cite{Banks}; the information about $U$ and $V$ is then
``stored'' in the relation to the original variables. (In this process
global information may be lost, as happens, e.g., if $U$ has critical
points. In the following we thus restrict our attention to such
choices of the potentials $U,V,W$ where the above transformation is
sufficiently global; otherwise the considerations to follow are of
local nature and the global information has to be restored in a
subsequent step, cf.\ also \cite{TK} for details.)  Following a recent
paper by Izquierdo \cite{Izq}, the explicit component form of the
supersymmetric extension of \re{gdil}), rewritten in Einstein-Cartan
variables, takes the form\footnote{After completion of this letter, I
  became aware that the action below was found already in
  \cite{Ikedasusy} (which was not noticed in \cite{Izq}, where the
  latter of the two works was cited in a different context only).
  \cite{Ikedasusy} contains also parts of the considerations to
  follow, but, such as in the case of bosonic Poisson
  $\sigma$--models, the hidden structure of a (graded) Poisson
  manifold is not noted (and, consequently, no statement about, e.g.,
  the classical solutions could be obtained  there).}  \widetext \be L =
\phi d\o + X_a (de^a + {\e^a}_b\omega e^b + 2i {\bar\psi}\gamma^a\psi
) - ( 2 uu' - \frac{iu''}{16} {\bar\chi}\chi ) \e_{ab} e^a e^b + 4i u
{\bar\psi}\gamma^3\psi + i u' {\bar\chi} e^a \gamma_a \psi + i
{\bar\chi} ( d\psi + \frac{1}{2}\omega \gamma_3 \psi ) \label{Izq} \,
. \ee \narrowtext Here $e^a$ is the zweibein of the metric $g$ (or the
conformally rescaled metric, respectively), ${\e^a}_b \, \o$ is the
spin connection, and $\psi$ is a oneform--valued Majorana spinor.
$\chi$ is a (zeroform--valued) Majorana fermion, which, such as
$\psi$, is of odd Grassmann parity, and $X^a$ denotes a pair of scalar
functions (the Lorentz index $a$ running over two values in two
spacetime dimensions). The fields $X^a$ are introduced in the bosonic
theory as Lagrange multipliers so as to determine $\o$ through the
vanishing torsion constraint; as the field equations for $\o$
determine $X^a$ uniquely in terms of the remaining variables, the
fields $X^a$ may be eliminated from (or introduced to) the action
without changing the theory (at least on the classical level).  $u$ is
a function of the dilaton $\phi$, which relates to $W$ in \re{gdil})
through $W = 4 (u^2)'$, the prime denoting differentiation with
respect to the argument $\phi$.  For the conventions we conform to
\cite{Izq}: the Lorentz metric has signature $(-,+)$, Lorentz indices
are underlined, $\e^{{\underline{0}}{\underline{1}}} = +1$, and the
gamma matrices are related to the usual Pauli matrices $\s_i$
according to: $\g^{\underline{0}} = -i \s_2$, $\g^{\underline{1}} =
\s_1$, and $\g^3 \equiv \g^{\underline{0}}\g^{\underline{1}}= \s_3$.
  
We now collect the one-- and zeroforms into two multiplets: \be
(X^i) := (X^a, \chi^\alpha, \phi) \; , \quad (A_i) := (e_a, i \bar
\psi_\a, \o) \label{multi} \; . \ee After a simple partial integration,
the action \re{Izq}) is seen to take the form \re{PSM}) and the
coefficient matrix $\CP^{ij}$ may be read off by straightforward
comparison:  \ba &\CP^{ab} = -\e^{ab} (4 u u' + \frac{1}{8i}
u'' \chi_\a \chi^\a ) \; , \quad \CP^{a \a } = u' (\gamma^a \chi)^\a
\; , & \nn &\CP^{\a\b} = - 8i u (\gamma^3)^{\a\b} - 4i X_a
(\gamma^a)^{\a\b} \, , & \label{PB} \\
& \{ X^a, \phi \} = \e^a{}_b X^b \; , \quad \{ \chi^\a, \phi \} =-\2
(\gamma^3 \chi)^\a & \nonumber \ea 
where $(\gamma^a \chi)^\a \equiv (\gamma^a)^\a{}_\b \chi^\b$, spinor
indices have been raised and lowered by means of $\e^{\a \b}$
($\chi_\a = \e_{\a \b} \chi^\b$ with $\e^{01}:=1$), and, in the last
line, the identification \re{bracket}) was used.
  
Up to now it may seem that we have not gained much and in particular
it is by no means clear that the matrix $\CP$ obtained above indeed
satisfies the (graded) Jacobi identity \re{Jacobi}), which, however,
is at the heart of Poisson $\s$--models.

In \cite{Izq} it was shown that the field equations of \re{Izq}) form
a free differential algebra (FDA) \cite{FDA}. In the present
formulation \re{PSM}) the field equations take the compact form \be
dX^i - A_j \CP^{ji} = 0 \,\, , \; \, \, dA_i - \2 A_k A_l (\CP^{lk}
\dl_{\! i}) =0 \,\, .\label{eom} \ee Applying an exterior
derivative to the first set of these equations, we obtain $dA_j \,
\CP^{ji} - A_j \, (\CP^{ji} \dl_{\! k}) \, dX^k=0$. Eliminating
$dA_j$ and $dX^k$ by means of \re{eom}), we end up with an expression
bilinear in the $A's$. By definition of an FDA, the resulting
equations have to be fulfilled identically, without any restriction to
the oneforms $A_i$. It is a simple exercise to show that this
requirement is fulfilled if and only if
$\CP^{ij}$ satisfies Eqs.\ \re{Jacobi}). Thus, using the result of
\cite{Izq}, the validity of the graded Jacobi identities is proven.

Certainly the Jacobi identities may be verified also by a direct
calculation using the specific form \re{PB}) of the bracket. This is
in fact {\em simpler\/} than proving the FDA property of the field
equations of the specific action \re{Izq}) (cf.\ also first sentence
of the following paragraph). Seeking brackets fulfilling the graded
Jacobi identities with restrictions specified below will {\em
  automatically}\/ provide 2d supergravity theories and to us this
route seems to be the technically simplest for the construction of
such models. This idea will be made clearer in what follows.

Applying an exterior derivative to the second set of equations, the
requirement for an FDA does not lead to any further relations beside
\re{Jacobi}).  Thus we may conclude that the field equations of a {\em
  general}\/ graded Poisson $\sigma$--model \re{PSM}) form an FDA, iff
the tensor $\CP$ is a Poisson tensor (i.e., by definition, iff $\CP$
satisfies \re{Jacobi})). Alternatively, the Jacobi identity may be
verified to be the necessary and sufficient condition for the
constraints in a Hamiltonian formulation of \re{PSM}) to be of first
class. This is tantamount to requiring the model to have maximal local
symmetries.

It is a nice and simple exercise to show that due to \re{Jacobi}) the
variations \be \delta X^i = \ep_j \CP^{ji} \; , \quad \delta A_i = d
\ep_i - A_j \ep_k (\CP^{kj} \dl_{\! i}) \label{symm} \ee change
the action only by a total divergence: $\delta S = \int d (\ep_i
dX^i)$. Thus there is a local symmetry for any pair of fields $(X^i,
A_i)$. Since the action is of first order in these fields, this
implies that there are at most a finite number of physical (gauge
invariant) degrees of freedom. (More precisely, being a oneform,
$A_i$ has two components for each value of $i$; however, the ``time
component'' $A_{0i}$ of the oneform $A_i$ enters a Hamiltonian
formulation as Lagrange multiplier for the constraints only and
therefore it must not be included in the above naive counting).

In the context of \re{Izq}) resp.\ \re{PB}) there are five independent
local symmetries contained in \re{symm}): $\ep_\phi$ generates local
Lorentz symmetries (cf.\ last line in Eq.\ \re{PB})).  The Grassmann
spinor $\ep_\a$, on the other hand, generates precisely the local
supersymmetry transformations of \cite{Izq}. The remaining two
symmetries correspond to the obvious diffeomorphism invariance of the
action. Indeed, a simple calculation shows that the Lie derivative of
$X^i$ and $A_i$ along a spacetime vector field $v$ differs from
\re{symm}) with the specific choice $(\epsilon_i) := (v^\m A_{\m i})$
only by an additive term proportional to the field equations
\re{eom}).  For invertible zweibein $A_a$ the first two entries
establish a bijection between any possible choice of $v$ and $\ep_a$,
moreover.  As this will be of some importance later on, we note,
however, that while diffeomorphisms cannot change an invertible matrix
$A_{\m a}$ into a noninvertible one and vice versa, the symmetries
\re{symm}) can.\footnote{Since the difference in the symmetries occurs
  only for nondegenerate metrics, which, however, in a gravitational
  theory are usually excluded, this difference is mostly irrelevant
  for what concerns the relevant factor spaces; cf., however,
  \cite{PLB} for counterexamples to this somewhat naive picture.}

We return to the structure found in the target space $N$ of the
theory. The target space is spanned by a Lorentz vector $X^a$, a
Majorana spinor $\chi^\a$, and the dilaton $\phi$. $X^a$ and $\chi^\a$
may be combined into a $(1+1)$--dimensional superspace. $\phi$, on the
other hand, generates Lorentz boosts in this superspace by means of
the Poisson brackets \re{PB}), last line. Indeed, $\e^a{}_b$ is the
Lie algebra element of the (one--dimensional) Lorentz group in the
fundamental representation and $\gamma^3$ is easily identified with
the generator of Lorentz boosts in a two--dimensional spinor space:
$\gamma^3 \equiv \g^{\underline{0}} \g^{\underline{1}} = \2
[\g^{\underline{0}}, \g^{\underline{1}}]$ (irrespective of the choice
of presentation for the generators $\g^{\underline{0}}$ and
$\g^{\underline{1}}$ in the Clifford algebra).

This structure in the target space will remain also within
generalizations to more general 2d supergravity theories including the
supersymmetrization of theories with nontrivial torsion, such as the
Katanaev--Volovich (KV) model \cite{KV}, for which a supersymmetrization
has not been provided in the literature yet. Using the same fields
\re{multi}) as before, such theories may be found by searching for
other solutions to the Jacobi identity \re{Jacobi}). However, the then
yet unknown Poisson tensor should be restricted to agree with the last
line in \re{PB}). This is due to the relation of the Poisson bracket
on the target space and the local symmetries \re{symm}) and thus
implicitly required by local Lorentz invariance, present in any
gravity theory when formulated in Einstein--Cartan variables.

It is worth noting that restricting the Poisson tensor only by the
last line of \re{PB}), the Jacobi identities \re{Jacobi}) with one of
the indices corresponding to $\phi$ requires the Poisson tensor
components $\CP^{ab}$, $\CP^{a \a}$, and $\CP^{\a\b}$ to transform
covariantly under Lorentz transformations! E.g.\ for $\CP^{a \a}$ the
Jacobi identities require $\{ \CP^{a \a}, \phi \} = \e^a{}_b\CP^{b\a}
-\2 (\g^3)^\a{}_\b \CP^{a\b}$.  Thus to obtain the most general
supergravity theory that fits into the present framework, we can
proceed as follows: It must be possible to build the unknown tensor
components of $\CP$ by means of the Lorentz covariant quantities
$X^a$, $\chi^\a$, $\e^{ab}$, $(\gamma^a)^{\a\b}$, and $(\g^3)^{\a\b}$
($\e^{\a\b}$ is incorporated automatically by raising and lowering
spinor indices) with coefficients that are Lorentz invariant
functions, i.e.\ functions of $X_aX^a$, $\chi_\a\chi^\a$, and $\phi$.
Thus, e.g., the antisymmetric tensor $\CP^{ab}$ {\em must}\/ be of the
form $\CP^{ab}= \e^{ab} (F_1 + \chi_\a \chi^\a F_2)$, where $F_{1,2}$
are functions of the two arguments $X_aX^a$ and $\phi$. The remaining
Jacobi identities then reduce to a (comparatively simple) set of
differential equations for these coefficient functions.

Proceeding in this way, e.g., by replacing all (five) coefficients in
its first two lines of the bracket \re{PB}) by yet undetermined
coefficient functions of $X^aX_a$ and $\phi$, one can show that the
remaining Jacobi identities \re{Jacobi}) force the coefficients to
agree with those provided already in \re{PB}) (except for a
simultaneous global prefactor). More general theories can thus be
obtained only by using further covariant entities to build $\CP^{a
  \a}$ and (possibly also) $\CP^{\a \b}$. Indeed, Lagrangians
quadratic in torsion require an extra additive term $F(X^aX_a,\phi)
X^a (\gamma^3)^\a{}_\b\chi^\b$ in $\CP^{a\a}$ \cite{Kummer}, also
perfectly compatible with Lorentz covariance. In \cite{prep} more
general supergravity theories will be constructed by the above method.
By construction, the resulting theories will be invariant against
superdiffeomorphisms incorporated within \re{symm}), thus allowing for
an interpretation as supergravity theory. The supersymmetrization of
the KV--model will be contained as a particular example in the class
of models constructed in this way.

\vspace{1em}

We finally turn to the solution of the field equations. Beside its
notational compactness the main advantage of the formulation \re{PSM})
as opposed to \re{Izq}) is its inherent target space covariance. Thus
we may change coordinates in the target space of the theory so as to
simplify the tensor $\CP$, while the field equations in the new
variables still will take the form \re{eom}) (but then with the
transformed, simplified Poisson matrix).  We will first use this
method to show that locally the space of solutions to the field
equations of \re{Izq}) modulo gauge symmetries is just
one--dimensional.  Thereafter we will provide a representative of this
one--parameter family in terms of the original variables used in
\re{Izq}). As a byproduct we will find that locally the space of
solutions is identical to the one of the bosonic theory; all the
fermionic fields may be put to zero by gauge transformations. Since
any global solution can be obtained by patching together local
solutions, we conclude that the local equivalence between the bosonic
theory \re{gdil}) and its supersymmetrization holds also on a global
level. It would be interesting, however, to confirm this result in a
more direct way.

Locally (more precisely, in the neighborhood of a generic point) any
(bosonic) Poisson manifold allows for Casimir--Darboux (CD)
coordinates $(C^A, Q^I, P_J)$ \cite{Weinstein}; constant values of the
Casimir functions $C^A$ label symplectic leaves in $N$, on which the
remaining coordinates are standard Darboux coordinates: $\{Q^I , P_J\}
= \d^I{}_J$ (all other brackets vanishing). According to
\cite{Henneaux}, Darboux coordinates exist also for supersymplectic
manifolds (manifolds with a graded, nondegenerate Poisson bracket). We
thus assume that CD coordinates exist in the case of general, graded
Poisson manifolds. However, at least in the case of the bracket
\re{PB}), they definitely do: Although we did not succeed to find such
coordinates explicitly in the present paper, we will provide a Casimir
function below. On its level surfaces (defined by the appropriate
quotient algebra of superfunctions) the Poisson bracket is (almost
everywhere) nondegenerate and the result on supersymplectic manifolds
may be applied.

We now are in the position to show that locally there is only a
oneparameter family of gauge inequivalent solutions to the field
equations of \re{Izq}). For this purpose we only need to know about
the local {\em existence}\/ of CD coordinates $\wt X^i$.  In these
coordinates\footnote{To be sure: These are coordinates on the target
  space, not on the worldsheet spacetime. From the point of view of
  the field theory a change of coordinates $X^i \to \wt X^i$, which
  induces the change $A_i \to \wt A_i \equiv A_j (\dl
  X^j/\partial \wt X^i)$, corresponds to a (local) change of field
  variables.}  the second set of field equations \re{eom}) reduce to
$d \wt A_i =0$. Thus locally $\wt A_i = d f_i$ for some functions
$f_i(x)$. However, the local symmetries \re{symm}) also simplify
dramatically in these new field variables: $\d \wt A_i = d \wt \e_i$.
This infinitesimal formula may be integrated easily showing that all
the functions $f_i$ may be put to zero identically.  But then we learn
from the first set of the field equations \re{eom}) that all the
functions $\wt X^i$ are constant. All of the constant values of $Q^I$
and $P_I$ may be put to an arbitrary value by means of the residual
gauge freedom in \re{symm}) (constant $\ep_i$). What remains as gauge
invariant information is only the constant values of the Casimir
functions $C^A$. Since the Poisson tensor \re{PB}) has rank four
(almost everywhere),\footnote{To determine the rank of the matrix
  $\CP^{ij}$, we may concentrate on the rank of the the two by two
  matrix $\CP^{\a\b}$ and the three by three matrix in the purely
  bosonic sector; $\CP^{a \a}$, being linear in Grassmann variables,
  cannot contribute to the rank of the matrix, cf.\cite{Henneaux}.}
the model defined by Eq.\ \re{Izq}) has just one Casimir function and
its space of local solutions is thus indeed one--dimensional only.

The local solution obtained above in terms of CD-coordinates may be
transformed back easily to any choice of target space coordinates. We
find that also in the original variables: $A_i \equiv 0$ and $X ^i =
const^i$, where the latter constants may be chosen at will as long as
they are compatible with the constant values of the Casimir(s) $C^A$,
which characterize the (local) solution. As it stands, this solution
corresponds to a solution with vanishing zweibein and metric. In a
gravitational theory, the metric (and zweibein) is required to be a
nondegenerate matrix, however. The vanishing zweibein is a result of
using the symmetries \re{symm}), which, in contrast to
diffeomorphisms, connect the degenerate with the nondegenerate sector
of the theory. A similar phenomenon occurs, e.g., also within the
Chern--Simons formulation of $(2+1)$--gravity. The problem may be
cured by applying a gauge transformation \re{symm}) to the local
solution $A_i \equiv 0$ so as to obtain a solution with nondegenerate
zweibein. However, in contrast to Chern--Simons theory, where the
behavior of $A$ under {\em finite}\/ (nonabelian) gauge
transformations is known, the infinitesimal gauge symmetries
\re{symm}) cannot be integrated in general (except for the case where
the Poisson tensor is (at most) linear in the fields $X^i$ and the theory
reduces to a (non)abelian gauge theory). We thus need to introduce one
further step. 

For the Poisson brackets \re{PB}) a possible choice for a Casimir
function $C$, $\{C, \cdot \} \equiv 0$, is \be C = \2 X_a X^a + 2 u^2 -
\frac{i}{8} u' \chi_\a \chi^\a \, . \label{C} \ee This expression
basically coincides with the Casimir found in the bosonic theory: as
there the last two terms are the integral of
$\CP^{{\underline{0}}{\underline{1}}}$ with respect to $\phi$,
$\chi_\a \chi^\a$ Poisson commuting with all bosonic variables. We now
choose new coordinates on (patches of) $N$ (where $X^a \neq 0$)
according to \be \bar X^i := (C,\ln |X^+|, \chi^\a, \phi) \, ,
\label{new} \ee with the null coordinates $X^\pm :=
(X^{\underline{1}}\mp X^{\underline{0}})/\sqrt{2}$.  (An argumentation
similar to the following one may be applied also if $\ln |X^+|$ is
replaced by $X^+$ in \re{new})). $(C,\ln |X^+|, \phi)$ provides a CD
coordinate system in the purely bosonic sector (cf.\ also \cite{TK});
however, in the five--dimensional target space $N$, these coordinates
are far from being CD, several brackets still containing the potential
$u(\phi)$. It thus seems rather difficult to solve the field equations
for field variables \re{new}). However, from our considerations above,
we know that up to Poisson--$\sigma$ gauge transformations the local
solution always takes the from $\bar A_i \equiv 0$, $C = const$, $\ln
|X^+|= \chi^\a = \phi =0$. In these field variables, induced by the
coordinates \re{new}), it is now possible to gauge transform this
explicitly to a solution with nondegenerate zweibein and, {\em
  simultaneously}, the resulting solution may be transformed back to
the original variables used in \re{Izq}) --- this is possible since in
contrast to the CD coordinates $\wt X^i$ used above, the coordinates
$\bar X^i$ are known explicitly in terms of the original variables.

It is straightforward to verify that {\em on the above solutions}\/
the infinitesimal gauge transformations \re{symm}) with $(\bar \ep_i)
:= (\ep_C, \bar \ep_+, 0, 0, 0)$ are simply: \be \delta A_C = \delta
\ep_C \, , \; \delta \bar A_+ = d \bar \ep_+ \, , \; \delta \phi =
\bar \ep_+ \, , \label{symm2} \ee with {\em all}\/ other variations
vanishing. Note that, e.g., in the second relation we dropped terms
proportional to $\bar A_\a$, since $\bar A_\a$ can be {\em kept}\/
zero consistently by the above transformations due to $\delta \bar
A_\a = 0$. It is thus possible to {\em integrate}\/ the gauge
symmetries \re{symm2}): $A_C \to A_C + df_1$, $\phi \to \phi + f_2$,
$\bar A_+ \to \bar A_+ + df_2$ ({\em all}\/ other fields remaining
unaltered) where $f_{1,2}$ is an arbitrary pair of functions on the 2d
spacetime. The degenerate solution is then transformed into $A_C =
df_1$, $\bar A_+ = df_2$, $\bar A_\alpha= \bar A_\phi = 0$, $C =
const$, $\ln |X^+|= \chi^\a =0$, and $\phi = f_2$. Using $f_1$ and
$f_2$ as coordinates $x^1$ and $x^0$ on the worldsheet, respectively,
and transforming these solutions back to the original variables
\re{multi}) (using $A_i = \bar A_j (\dl \bar X^j/\partial X^i)$), we
obtain: \ba &(e^+,e^-,\o) = (dx^1 , dx^0 + \2 h(x^0) dx^1,- h'(x^0)
dx^1) \, ,& \; \nn &(X^+,X^-,\phi) = (1 , \2 h(x^0) , x^0) \, , &
\label{Ende} \ea where $h(x^0) \equiv C - 2 u^2(x^0)$, $C$ being the
constant value of the Casimir \re{C}).  {\em All}\/ the fermionic
variables vanish identically.

Thus, up to gauge transformations, the local solution agrees
completely with the one found in the purely bosonic dilaton theory
\re{gdil}). This applies at least to those patches where the above
coordinate systems are applicable. Since, e.g., all the fixed points
of the supersymmetric bracket \re{PB}) lie entirely within the bosonic
sector of the target space, we expect that there are also no
exceptional solutions, containing (necessarily) nonvanishing fermionic
fields. Moreover, the subsequent global analysis of \cite{TK} may be
applied to the solutions \re{Ende}) without change.

So the characterization of the dynamics of the general supersymmetric
extension of \re{gdil}) turns out to be less difficult than it appeared
at the time when \cite{Strominger} was written (cf the remarks following
equation (50) of that paper). Rather, it appears that the
supersymmetric extension of \re{gdil}) is {\em trivial (on--shell)\/},
at least at the classical level.

It would be interesting to check this result by some other method and
to possibly establish it in a less indirect way. It is to be expected,
moreover, that a similar result holds also on the quantum level.

 Let us finally remark that the
supersymmetric extension may still be of some value even on the purely
classical level: In \cite{Strominger} it was used, e.g., to establish
the positivity of (some notion of) the ``mass'' (presumably closely
related to the Casimir $C$ above, cf.\ \cite{Lau}) in a large class of
(nonsupersymmetric) models \re{gdil}) {\em coupled}\/ to matter
fields.  Further investigations of 2d dilatonic supergravity theories,
including generalizations to theories with nontrivial torsion and a
comparison to the existing literature \cite{Ertl} is in preparation
\cite{prep}.

\begin{acknowledgements} The author  gratefully acknowledges
  discussions with M.\ Ertl, C.\ Gutsfeld, W.\ Kummer, D.\ Roytenberg
  and P.\ Schaller.
\end{acknowledgements}

\widetext


\begin{thebibliography}{99}
\bibitem{Banks} T.\ Banks and M.\ 
  O'Loughlin, {\em Nucl.\ Phys.}\/ {\bf B362} (1991), 649; S.D.\ 
  Odintsov, I.L.\ Shapiro, {\em Phys.\ Lett.}\/ {\bf B263} (1991),
  183;  D.\ Louis-Martinez, J.\ Gegenberger, G.\ 
   Kunstatter, {\em Phys. Lett.}\/ {\bf B321} (1994), 193.
\bibitem{Strominger} Y.\ Park and A.\ Strominger, {\em Phys.\ Rev.}
  {\bf D47}, 1569 (1993). 
\bibitem{Howe} P.S.\ Howe, {\em J.\ Phys.}\/ {\bf A12}, 393 (1997). 
\bibitem{PSM} P.\ Schaller and T.\ Strobl, {\em Mod.\ Phys.\ Lett.}\/
  {\bf A9} (1994), 3129; T.\ Strobl, {\em Poisson Structure Induced
    Field Theories and Models of 1+1 Dimensional Gravity}, PhD thesis,
  Vienna 1994; P.\ Schaller and T.\ Strobl, hep-th/9411163; A.\ 
  Alekseev, P.\ Schaller and T.\ Strobl {\em Phys.\ Rev.}  D 52 (1995)
  7146. For older related work cf.\ N.\ Ikeda, {\em Ann.\ Phys.}\,
  {\bf 235}, 435 (1994) and P.\ Schaller and T.\ Strobl, {\em Lecture
    Notes in Physics}\/ {\bf 436} (1994) 98, Eds.\ A.\ Alekseev et al,
  Springer. For some recent progress cf.\ A.S.\ Cattaneo and G.\ 
  Felder, {\em A Path Integral Approach to the Kontsevich Quantization
    Formula}\, math/9902090.  
\bibitem{LNP} P.\ Schaller and T.\ 
  Strobl, {\em Lecture Notes in Physics} 469, p.\ 321-333,
  'Low-Dimensional Models in Statistical Physics and Quantum Field
  Theory', Eds.\ H.\ Grosse and L.\ Pittner, Springer 1996
  (hep-th/9507020). 
\bibitem{Henneaux}M.\ Henneaux and C.\ 
  Teitelboim, {\em Quantization of Gauge Systems}, Princeton
  University Press 1992.  
\bibitem{TK} T.\ Kl\"osch and T.\ Strobl,
  {\em Class.\ Quantum Grav.}\/ {\bf 13} (1996) 965, Corr.\ {\bf 14}
  (1997), 825; {\em ibid.}\/ {\bf 13} (1996) 2395; {\em ibid.}\/ {\bf
    14} (1997), 1689.  
\bibitem{Izq}J.M.\ Izquierdo, {\em Phys.\ Rev.\/} {\bf D59} (1999)
  084017. 
\bibitem{Ikedasusy} 
N.\ Ikeda, {\em Int.\ J.\ Mod.\ Phys.\/} {\bf A9} (1994)  
1137;  {\em Ann.\ Phys.\/} {\bf 235} (1994) 435.
\bibitem{FDA} D.\ Sullivan, {\em Inst.\ des Haut
    \'Etud.\ Sci.\ Pub.\ Math.}\/ {\bf 47}, 269 (1977).
\bibitem{PLB} P.\ Schaller and T.\ Strobl, {\em Phys.\ Letts.}\/  
{\bf B337}, 266  (1994);  H.J.\ Matschull, {\em On the relation
  between 2+1 Einstein gravity and Chern Simons theory},
gr-qc/9903040. 
\bibitem{KV} M.O.\ Katanaev and I.V.\ Volovich, {\em Phys.\ Lett.}\/
  {\bf B175} (1986) 413; M.O.\ Katanaev, {em J.\ Math.\ Phys.}\/ {\bf
    31}, 882 (1990); W.\ Kummer and D.J.\ Schwarz, {\em Phys.\ Rev.}\/
  {\bf D45}, 3628 (1992); P.\ Schaller and T.\ Strobl, {\em Class.\ 
    Quant.\ Grav.} {\bf 11} (1994) 331.
\bibitem{Kummer} W.\ Kummer, private communication. 
\bibitem{prep}  M.\ Ertl, W.\ Kummer and T.\ Strobl, in preparation.
\bibitem{Weinstein} A.\ Weinstein, {\em J.\ Diff.\ Geom.} {\bf 18}, 523
  (1983). 
\bibitem{Lau} W.\ Kummer and S.\ Lau,
  {\em Annals Phys.}\/ {\bf 258} (1997) 37-80; J.\ Gegenberg, G.\ 
  Kunstatter and D.\ Louis--Martinez, {\em Phys.\ Rev.} {\bf D57},
  3537 (1998).  
\bibitem{Ertl}  S.\ Nojiri and I.\ Oda, {\em Mod.\ Phys.\
    Lett.\/} {\bf A8} (1993) 53. A.\ Bilal, {\em Phys.\ Rev.\/} 
    {\bf D48} (1993) 1665. A.H.\ Chamseddine, {\em Phys.\ Lett.} {\bf B258},
  97 (1991). D.\ Cangemi and M.\ Leblanc, {\em Nucl.\ Phys.}  {\bf
    B420}, 363 (1994). V.O.\ Rivelles, {\em Phys.\ Lett.\/} {\bf B321}
  (1994) 189. M.\ Ertl, M.O.\ Katanaev and W.\ 
  Kummer, {\em Nucl.\ Phys.}\/  {\bf B530}, 457 (1998).
  
\end{thebibliography}
\end{document}